\documentclass[pra, manuscript,showpacs,superscriptaddress]{revtex4}
\usepackage{amsmath}
\usepackage{amsfonts}
\usepackage{amssymb}
\usepackage{amsthm}
\usepackage{graphicx}
\usepackage{bm}
\usepackage{enumerate}

\begin{document}
\theoremstyle{plain} \newtheorem{theorem}{Theorem} \newtheorem{lemma}[theorem]{Lemma} \newtheorem{corollary}[theorem]{Corollary} \newtheorem{proposition}[theorem]{Proposition} \newtheorem{conjecture}[theorem]{Conjecture}

\theoremstyle{definition} \newtheorem{definition}{Definition}

\theoremstyle{remark} \newtheorem*{remark}{Remark} \newtheorem{example}{Example}

\title{Properties of the largest fragment in multifragmentation: a canonical 
thermodynamic calculation}
 
\author{G. Chaudhuri \footnote{On leave from Variable Energy Cyclotron Center, 1/AF Bidhan Nagar, Kolkata 700064, India}}
\email{gargi@physics.mcgill.ca}
\affiliation{Physics Department, McGill University, 
Montr{\'e}al, Canada H3A 2T8}

\author{S. Das Gupta}
\email{dasgupta@physics.mcgill.ca}
\affiliation{Physics Department, McGill University, 
Montr{\'e}al, Canada H3A 2T8}

\date{\today}

\begin{abstract}
Many calculations for the production of light and intermediate particles 
resulting from heavy ion collisions at intermediate energies exist.
Calculations of properties of the largest fragment resulting from
multifragmentation are rare.  In this paper we compute these properties
and compare them with the data for the case of gold on carbon.  We
use the canonical thermodynamic model.  The model also gives a bimodal
distribution for the largest fragment in a narrow energy range.

\end{abstract}

\maketitle

\section{Introduction}
The statistical model of nuclear disassembly in heavy ion collisions
is quite successful.  Here one assumes that the disintegrating system
with a given excitation energy expands to greater than normal volume 
before it breaks up into many composites of various sizes.  
Interactions between different composites in the rarefied situation
can be neglected and the
break up can be calculated using the laws of equilibrium thermodynamics.
This basic assumption is implemented in different versions according
to degrees of sophistication and detail.  Thus we have the statistical
multifragmentation model (SMM) of Copenhagen \cite{Bondorf}, the
microcanonical model of Gross \cite{Gross} and Randrup and Koonin
\cite{Randrup}.  An easily implementable canonical ensemble model
was later introduced \cite{Dasgupta1}.  The details of the model
and many applications can be found in a recent publication \cite{Das}.

It has been customary to pay a great deal of attention to the production
cross-sections of intermediate mass fragments and light charged
particles.  Here we study the properties of the largest fragment that
emerges in multifragmentation.  These properties are not easy to study
but the canonical model allows for such computation.  As we will see,
the heaviest fragment also reveals some interesting physics.  Data
on the heaviest fragment and fragments with the maximum charge
were published by EOS collaboration \cite{EOS1,EOS2,EOS3}.  In these 
experiments one studied the disintegrations of projectile-like
excited fragments in the reactions Au+C, La+C and Kr+C.  We do our
calculations for Au+C.  Calculations for the other two systems will
be similar.

In section II we write down the theoretical formulae needed for the 
calculations.  Results are presented in section III.  Summary and conclusions
are presented in section IV.

\section{Formulae in the canonical model}
The formulae used in this calculation are provided here.  Simpler
formulae for a hypothetical system of one kind of particles are given in 
\cite{Das}.

We will consider disassembly of the projectile-like fragment(PLF) where the
PLF is formed by the shearing off of a part of
$^{197}$Au (by the $^{12}$C target).  This PLF will have a charge $Z$ 
(usually less than 79) and a neutron number $N$ 
(usually less than 118).  This PLF will break up into many composites with
charges $i$ and neutron number $j$ (for example, $^9$Be has $i=4$
and $j=5$).  If the number of composite with proton and neutron numbers
$i$ and $j$ respectively is $n_{i,j}$ then conservation of charge and 
baryon number
dictates that $Z=\sum i\times n_{i,j}$ and $N=\sum j\times n_{i,j}$.

The canonical partition function for the system at a given temperature $T$
is given in our model by
\begin{eqnarray}
Q_{Z,N}=\sum\prod \frac{\omega_{i,j}^{n_{i,j}}}{n_{i,j}!}
\end{eqnarray}
Here the sum is over all possible channels of break-up (the number of such
channels is enormous) which satisfies the conservation laws; $\omega_{i,j}$ 
is the partition function of one composite with
proton number $i$ and neutron number $j$ respectively and $n_{i,j}$ is
the number of this composite in the given channel.  This is a low-density/
high temperature approximation the justification of which is
demonstrated in \cite{Jennings} and \cite{Das}.  For a given channel
the sum $\sum_{i,j}n_{i,j}$ is called the multiplicity of the channel.
The one-body partition
function $\omega_{i,j}$ is a product of two parts: one arising from
the translational motion of the composite and another from the
intrinsic partition function of the composite:
\begin{eqnarray}
\omega_{i,j}=\frac{V_f}{h^3}(2\pi m(i+j)T)^{3/2}\times z_{i,j}(int)
\end{eqnarray}
Here $m(i+j)$ is the mass of the composite and
$V_f$ is the volume available for translational motion; $V_f$ will
be less than $V$, the volume to which the system has expanded at
break up. We use $V_f = V - V_0$ , where $V_0$ is the normal volume of the PLF.
 We will shortly discuss the choice of $z_{i,j}(int)$
used in this work.

The probability of a given channel $P(\vec n_{i,j})\equiv P(n_{0,1},
n_{1,0},n_{1,1}......n_{i,j}.......)$ is given by
\begin{eqnarray}
P(\vec n_{i,j})=\frac{1}{Q_{Z,N}}\prod\frac{\omega_{i,j}^{n_{i,j}}}
{n_{i,j}!}
\end{eqnarray}
The average number of composites with $i$ protons and $j$ neutrons is
seen easily from the above equation to be
\begin{eqnarray}
\langle n_{i,j}\rangle=\omega_{i,j}\frac{Q_{Z-i,N-j}}{Q_{Z,N}}
\end{eqnarray}
The constraints $Z=\sum i\times n_{i,j}$ and $N=\sum j\times n_{i,j}$
can be used to obtain different looking but equivalent recursion relations
for partition functions.  For example
\begin{eqnarray}
Q_{Z,N}=\frac{1}{Z}\sum_{i,j}i\omega_{i,j}Q_{Z-i,N-j}
\end{eqnarray}
Instead of labelling partition functions by their charge and neutron numbers
$Z$ and $N$ we can for example label them by total mass ($A\equiv Z+N$)
and charge $Z$.  Labelling them by $\tilde Q$ we have
\begin{eqnarray}
\tilde Q_{A,Z}=\frac{1}{A}\sum_{a,i} a\tilde\omega_{a,i}\tilde Q_{A-a,Z-i}
\end{eqnarray}  
These recursion relations allow one to calculate $Q_{Z,N}$ 
(equivalently $\tilde Q_{A,Z}$) very quickly in the computer.

We list now the properties of the composites used in this work.  The
proton and the neutron are fundamental building blocks 
thus $z_{1,0}(int)=z_{0,1}(int)=2$ 
where 2 takes care of the spin degeneracy.  For
deuteron, triton, $^3$He and $^4$He we use $z_{i,j}(int)=(2s_{i,j}+1)\exp(-
\beta e_{i,j}(gr))$ where $\beta=1/T, e_{i,j}(gr)$ is the ground state energy
of the composite and $(2s_{i,j}+1)$ is the experimental spin degeneracy
of the ground state.  Excited states for these very low mass
nuclei are not included.  For mass number $A=5$ and greater we use
the liquid-drop formula.  For nuclei in isolation, this reads ($a=i+j$)
\begin{eqnarray}
z_{i,j}(int)=\exp\frac{1}{T}[W_0a-\sigma(T)a^{2/3}-\kappa\frac{i^2}{a^{1/3}}
-s\frac{(i-j)^2}{a}+\frac{T^2a}{\epsilon_0}]
\end{eqnarray}
The derivation of this equation is given in several places \cite{Bondorf,Das}
so we will not repeat the arguments here.  The expression includes the 
volume energy, the temperature dependent surface energy, the Coulomb
energy, the symmetry energy and contribution from excited states
since the composites are at a non-zero temperature.  For $i,j$, 
(the proton and the neutron number)
we include a ridge along the line of stability.  We have used two
sets to test the sensitivity of the results to the width of
of the ridge.  We call these set 1 and set 2.  In set 1
for each $a$ between 5 and 40, 5 isotopes
are included (case for $a$=4 and lower was already mentioned before);
For $a>40$, 7 isotopes are included for each $a$.  In set 2, 5 isotopes
are used for $a$ between 5 and 9, 7 isotopes are used for $a$ between
10 and 40 and 9 isotopes for each $a>40$.  The results are quite
similar in most cases except for one figure which we will point out
when we compare with data.  It should be pointed out that enlarging
the width of the ridge does not necessarily imply a better calculation
as one may begin to overcount the phase space.

The Coulomb interaction is long range.  The Coulomb interaction between
different composites can be included in an approximation called
the Wigner-Seitz approximation.  We incorporate this following the
scheme set up in \cite{Bondorf}.  This requires adding in the argument
of the exponential of Eq.(7) a term 
$\kappa\frac{i^2}{a^{1/3}}(V_0/V)^{1/3}(\frac{Z_0/A_0}{i/a})^{1/3}$.  Here
$A_0, Z_0$ are the mass and charge number respectively
of the disintegrating system, $V_0$ is the
normal nuclear volume for this system and $V$, the freeze-out volume
(typically 4-5 times $V_0$).  Defining $R_f^3\equiv\frac{3V}{4\pi}$
the energy of the system is given by 
$E=\frac{3Z_0^2}{5R_f}+\sum_{i,j}\langle n_{i,j}\rangle e_{i,j}$
where for $a>4$ we have
$e_{i,j}=\frac{3}{2}T+a(-W_0+T^2/\epsilon_0)+\sigma(T)a^{2/3}+
\kappa\frac{i^2}{a^{1/3}}[1.0-(V_0/V)^{1/3}
(\frac{Z_0/A_0}{i/a})^{1/3}]+s\frac{(i-j)^2}
{a^2}-T[\partial\sigma(T)/\partial T]a^{2/3}$.  For $a\leq 4$ we use
$e_{i,j}=\frac{3}{2}T+e_{i,j}(gr)-
\kappa\frac{i^2}{a^{1/3}}(V_0/V)^{1/3}(\frac{Z_0/A_0}{i/a})^{1/3}$.  We label
as $E^*$ the excitation energy: $E^*=E-E(gr)$ where $E(gr)$ is 
calculated for mass number $A_0$ and charge $Z_0$ using the liquid-drop
formula.

The central issue in this work is the calculation of $\langle Z_{max}\rangle$
and $\langle A_{max}\rangle$ (and their fluctuations) when the fragmenting 
system has charge $Z_0$ and mass number $A_0$.  Let us fix on 
calculating $\langle Z_{max}\rangle$.  The calculation for 
$\langle A_{max}\rangle$ can be done by analogy with appropriate and
obvious changes in the formulae.  There is an enormous number of
channels in Eq.(1).  Different channels will have different values 
of $Z_{max}$.  For example there is a term $\frac{\omega_{1,0}^Z}{Z!}
\frac{\omega_{0,1}^N}{N!}$ 
in the sum of Eq.(1).  In this channel $Z_{max}$ is 1.  The probability
of this channel occurring is (from Eq.(3)) 
$\frac{1}{Q_{Z,N}}\frac{\omega_{1,0}^Z}{Z!}\frac{\omega_{0,1}^N}{N!}$.
The full partition function can be written as
$Q_{Z,N}=Q_{Z,N}(\omega_{1,0},\omega_{0,1},
\omega_{1,1},........\omega_{i,j}......)$.  If we construct a $Q_{Z,N}$
where we set all $\omega$'s except $\omega_{1,0}$ and $\omega_{0,1}$
to be zero then this 
$Q_{Z,N}(\omega_{1,0},\omega_{0,1},0,0,0..............)
=\frac{\omega_{1,0}^Z}{Z!}\frac{\omega_{0,1}^N}{N!}$ and this has
$Z_{max}=1$.  Consider now constructing a $Q(Z,N)$ with only three
$\omega$'s: $Q_{Z,N}(\omega_{1,0},\omega_{0,1},\omega_{2,1},0,0,0,......)$.
This will have $Z_{max}$ sometimes 1 
(as $\frac{\omega_{1,0}^Z}{Z!}\frac{\omega_{0,1}^N}{N!}$ is still there)
and sometimes 2 (as, for example, in the term
$\frac{\omega_{2,1}^3}{3!}\frac{\omega_{1,0}^{Z-6}}{(Z-6)!}\frac{\omega_{0,1}^{N-3}}{N!}$). 

We are now ready to write down a general formula.  Let us ask the question:
what is the probability that a given value $z_m$ occurs as the maximum
charge?  To obtain this we construct a $Q_{Z,N}$ where we set all values of
$\omega_{z,j}=0$ when $z>z_m$.  Call this $Q_{Z,N}(z_m)$.  Then 
$Q_{Z,N}(z_m)/Q_{Z,N}$ (where $Q_{Z,N}$ is the full partition function with
all the $\omega$'s) is the probability that the maximum charge is
any value between 1 and $z_m$.  Similarly we construct a $Q_{Z,N}(z_m-1)$
where $\omega_{z,j}$ is set at zero whenever $z>z_m-1$.  The probability $p$
that $z_m$ is $Z_{max}$ is given by 
\begin{eqnarray}
p(z_m)\equiv p(Z_{max}=z_m)=\frac{Q_{Z,N}(z_m)-Q_{Z,N}(z_m-1)}{Q_{Z,N}}
\end{eqnarray}
The average value of $Z_{max}$ at given temperature and for given $Z_0,A_0$
is
\begin{eqnarray}
Z_{max}=\sum_{z_m=1}^{z_m=Z_0}z_mp(z_m)
\end{eqnarray}
and the fluctuation is
\begin{eqnarray}
RMS(Z_{max})=\sqrt{\sum p(z_m)(z_m-Z_{max})^2}
\end{eqnarray} 
Before we end this section we want to mention that the largest composite
we obtain at the end of the above calculation is at a finite temperature
$T$ and can further decay by evaporation, ending up in a lower mass
or charge number.  From our past experience \cite{Das} we know what
the effect will be: $Z_{max}$ will decrease slightly but the effect on
the $RMS(Z_{max})$ will be significant.  Without the evaporation the calculated
$RMS$ will be overestimated (see Fig.19 of \cite{Das}).  Inclusion of 
evaporation as an after burner is not possible at this stage.  The sum over
$z_m$ (Eq.(9)) is too huge and for each $z_m$ there is a sum over neutron 
number.  We will hope to get $Z_{max}$ nearly right but will overestimate
the $RMS$ value.  The next section compares data with our calculations.

\section{Decay of excited projectile-like fragmant}
In the EOS experiment part of the projectile (Au, La and Kr) is sheared off
by the $^{12}$C target.  We will compare our calculations with the Au+C data.
The size of the excited PLF which decays depends upon the impact parameter
which also determines the amount of excitation energy per nucleon
in the excited PLF.  In Figs.1(b) and 1(c) of \cite{EOS1} 
the size of the excited PLF 
(the size and charge are denoted by $A_0$ and $Z_0$) is plotted 
as a function of excitation energy $E^*$ per nucleon.  This aspect of the
experiment depends upon dynamics and is outside the scope of
a thermodynamic model.  However, given $E^*$ and $A_0,Z_0$, this excited PLF
will expand and break up into many pieces and this is calculable in a 
canonical thermodynamic model.
In our Fig.1 (data taken from Fig.1(f) of \cite{EOS1})
we have plotted $Z_{max}/Z_0$ as a function of $E^*$ per nucleon 
where $Z_{max}$ is the average maximum charge carried by a composite.  The data
are shown as points joined by a dotted curve.  The other curves are 
exploratory calculations.  Our fit to experimental data is given in  
Fig.2.  We now explain how the calculations are done.

For a given value of $A_0,Z_0$, the experiment provides the value of $E^*$.
The beginning of a canonical thermodynanic model besides $A_0$, $Z_0$, are 
 a temperature and a freeze-out density $\rho/\rho_0$ 
(freeze-out density in unit of normal nuclear density $\rho_0$=
0.16fm$^{-3}$) which
will then provide all the observables including $E^*$.  For a fixed $E^*$ the
canonical model employs a fixed $\rho/\rho_0$.  For example,
for central collisions of Sn on Sn at 50 Mev per nucleon beam energy,
a freeze out density of one-sixth normal density gives good results
for production cross-sections of intermediate mass fragments
\cite{Das}.  However, in the
EOS experiment $E^*$ varies over a wide range (from less than 2 MeV per nucleon
where the validity of the thermodynamic model as used here can be questioned
to 10 Mev per nucleon where the thermodynamic model is expected to work well)
and hence we should expect that $\rho/\rho_0$ will also need to vary in this
interval.  In general, the freeze-out density will decrease as $M/A_0$
($M$=multiplicity, $A_0$=mass number of the dissociating system)
increases, reaching some asymptotic value for large multiplicity.  
In SMM \cite{Bondorf}
the freeze-out density varies in each channel, decreasing as the multiplicity
increases (this makes Monte-Carlo simulation mandatory).  In the
canonical model, at a given temperature, the freeze-out density is kept
fixed irrespective of channels.  Thus the freeze-out density can be 
dependent only on the average multiplicity.  Past comparisons with SMM
predictions showed that at least for the obsrvables studied so far
this simplification is quite adequate \cite{Tsang}.

Fig.1 compares the data if in the calculation 
the freeze-out volume is kept
fixed at $\rho/\rho_0$ =0.25 (a typical canonical model value) or at 0.39
(this is an often quoted value in the model used in \cite{EOS1,EOS2}).
The value 0.39 is clearly better at low values of $E^*$ but leaves
too large a residue at higher $E^*$ whereas the value 0.25 is better
at the higher end of $E^*$ but is an underestimation at the lower end
of $E^*$.  The data undoubtedly point to the need of a variable $\rho/\rho_0$
if the whole spectrum of $E^*$ is to be covered.  In Fig.1, for brevity
we show results with set 1 only (see previous section for the range of nuclei
covered in set 1).
Fig.2 compares data with our calculation where we use a variable 
$\rho/\rho_0$.
We have used a parametrisation $\rho/\rho_0=a+b\exp(-c(E^*/A_0))$ where
$a$=0.17, $b=0.83$ and $c=0.417$ MeV$^{-1}$.  No optimisation of the fit
was tried but the values are suitable for the low and high limits of $E^*$.
Our values of
$\rho/\rho_0$ are also very similar to those quoted in 
Table II of \cite{EOS3} where a different model was used.  We have
shown results with set 1 and set 2 (this includes a larger number of composites
than set 1; see previous section).

In the canonical model calculation normally the inputs are the freeze-out
density and temperature.  Here we use the freeze-out density and $E^*$.
Given this density and $E^*$ we find the temperature which would give back
the $E^*$ we started with.  We then calculate $Z_{max}$.  Calculations
for $A_{max}$ were also done and the fits are similar.

Next we turn to results for $\gamma_2$ and $RMS(Z_{max}/Z_{0})$ which are shown
in Fig.3.  Data are from Fig.2 of \cite{EOS1}. Here
$\gamma_2=\frac{M_2M_0}{M_1^2}$ where $M_k$ is the $k$th moment of the
fragment distribution: $M_k=\sum n_aa^k$.  The over-estimation in 
the calculation for $RMS$ was already alluded to in the previous section.
Afterburner for the largest fragment will bring the distribution more
closely packed near the line of maximum stability thus reducing the
fluctuation.

The probability distribution (Eq.8) of $Z_{max}$ as a function of
$Z_{max}$ for EOS experiments is not known to
us but in view of the recent interest in such distributions 
\cite {Gulminelli,Pichon} we have computed the distributions and displayed
them in Fig.4.  We find that there is a window where the distribution
is bimodal.  Usually for discussion of bimodality one uses a fixed
freeze-out volume and varies the temperature but here, as we
discussed before, our freeze-out density actually changes as $E^*$
(hence $T$) changes but the nature of bimodality is still clearly
seen (if we fix the freeze-out density the plots are very similar).
Probability distribution for $A_{max}$ (calculated but not
shown here) also has the window for bimodality.  Connection between bimodality
and first order phase transition in the canonical model (which does have
a first order phase transition) is being pursued and we also hope 
to do calculations for other experimental data.

\section{Summary and Discussion}
The canonical thermodynamic model clearly reproduces many important
features of the largest fragment resulting from multifragmentation.
The model is very easy to implement and yet is very realistic.
The same model gives remarkable fits to light charged particles,
intermediate mass fragments and as we have just seen, the largest
fragment as well. We have thus far reported the calculation of properties of the largest fragment
without making any link with the subject of liquid-gas phase transition.  We
make a brief connection here.  In a large
system(see details in a recent article \cite{Chaudhuri}) composites with charge
$z\leq 20$ comprise the gas phase.  At co-existence, there will be, in addition,
one large composite which is the liquid.  This picture gets blurred as the
system size is reduced, but , nonetheless, the largest fragment (at
temperatures below that which displays bimodality) is an approximate scaled down version of the
liquid.  In a finite system bimodality in the largest fragment
distribution occurs in the temperature (energy) window where the system passes
from the liquid-gas co-existence phase to the pure gas
phase. The largest cluster has been discussed in the literature before \cite{Gulminelli, Krishnamachari, Pleimling, Binder}. However these all use the lattice gas model or
the Ising model (often with fixed magnetisation).  Our approach
here is more directly related to nuclear physics phenomenology
with phase transition issues in the background.

\section{Acknowledgement} 
This work is supported by the Natural Sciences and Engineering Research 
Council of Canada.

\pagebreak

\begin{figure}[htb]
\begin{center}
{\includegraphics[angle=0,width=15cm]{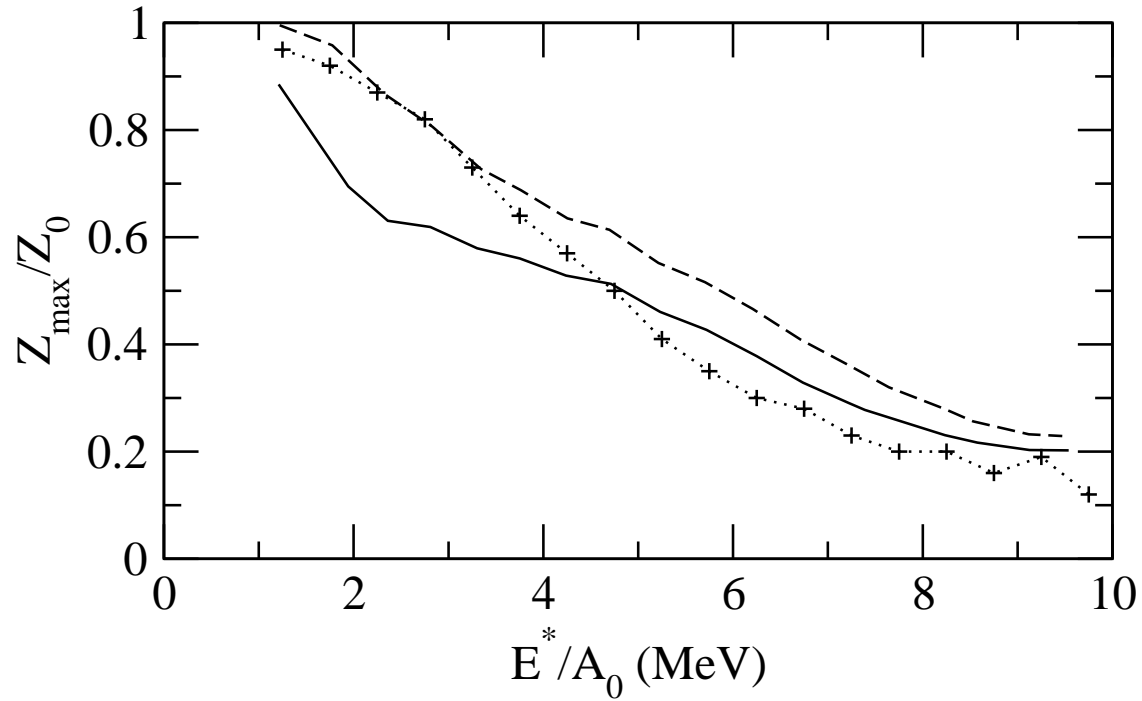}}
\end{center}
\caption{Experimental points (from Fig.1(f) of \cite{EOS1} ) are 
joined
by a dotted line to guide the eye.  The calculations are exploratory:
these are done for
a fixed freeze-out density (solid line for $\rho/\rho_0$=0.25 and
dashed line for $\rho/\rho_0$=0.39). A 
higher freeze-out density produces a higher value of $\langle Z_{max}
\rangle$.  In the figures $Z_{max}$ stands for this average value.
The calculations shown here
include the composites of set 1.  See the discussion following
eq.(7) for enumeration of composites included in set 1 and set 2.}

\end{figure}

\pagebreak

\begin{figure}[htb]
\begin{center}
{\includegraphics[angle=0,width=15cm]{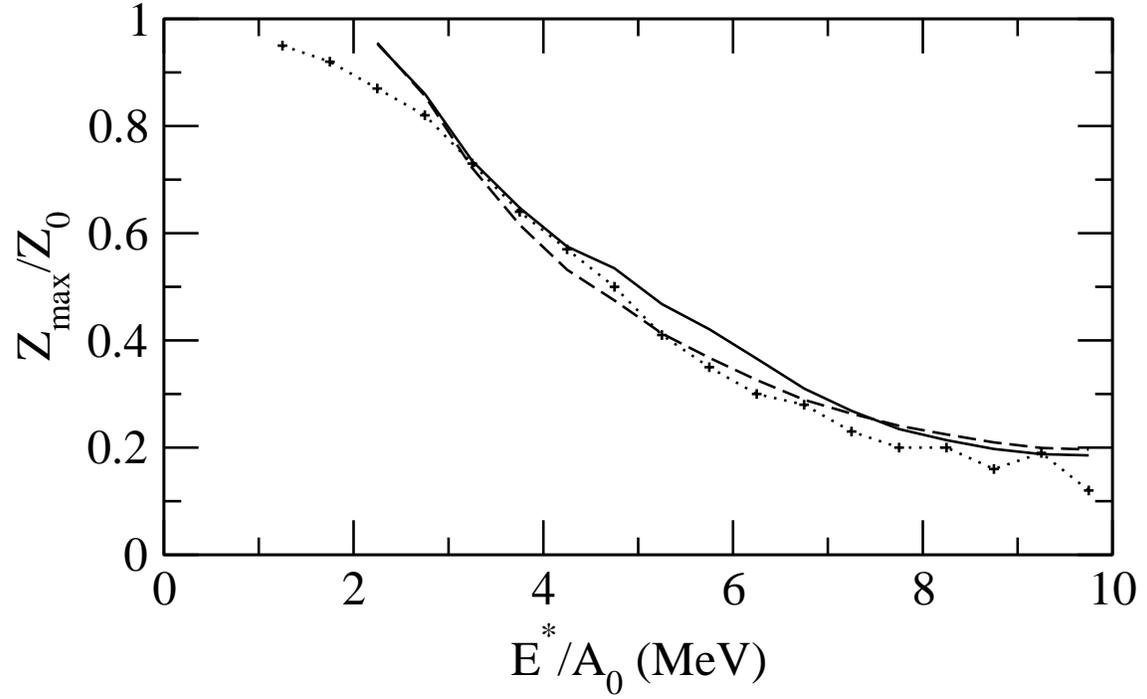}}
\end{center}
\caption{It is expected that the freeze-out density will decrease
with excitation energy in this energy range.  The calculations here
use a parametrisation $\rho/\rho_0=a+b\exp (-c(E^*/A_0))$ where
$a$=0.17, $b$=0.83 and $c$=0.417 MeV$^{-1}$.  The solid line uses
set 1, the dashed line uses set 2.  Experimental data are joined
by a dotted line.}

\end{figure}

\pagebreak

\begin{figure}[htb]
\begin{center}
{\includegraphics[angle=0,width=10cm]{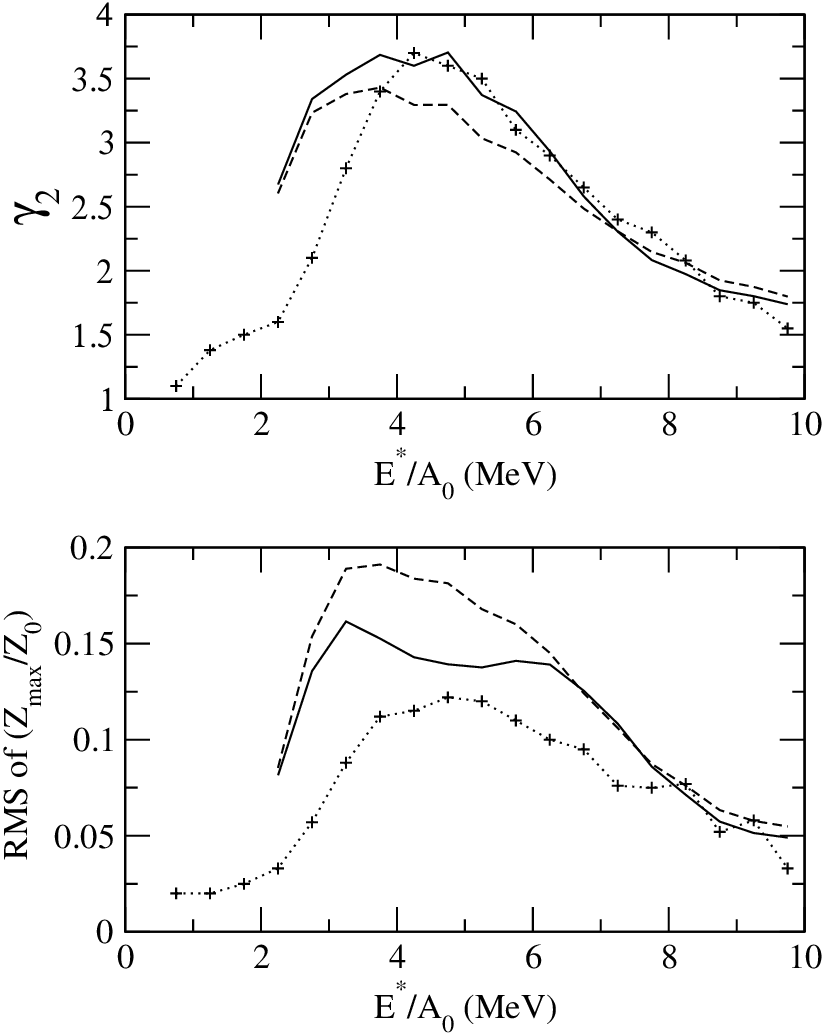}}
\end{center}
\caption{Top panel compares experimental data for $\gamma_2$
(fig.2 of \cite{EOS1}) with calculations.  For definition of
$\gamma_2$, see the text.  The solid curve uses set 1 and the dashed
curve uses set 2.  The bottom panel compares experimantal values of
$RMS$ of $Z_{max}$ with the canonical calculation.  As explained in the 
text,
without an afterburner evaporation calculation, $RMS$ values from 
theory
will significantly overestimate actual $RMS$.  
Evaporation will bring the population 
much closer to the line of maximum stability.  For an estimate of 
effects
of evaporation see fig.19 of \cite{Das}.  It is not surprising that 
with
set 2 (dashed curve) calculated values of $RMS$ exceed those calculated
with set 1.  As more nuclei further from the line of maximum stability
are included the effects of evaporation will be stronger.}

\end{figure}

\pagebreak
\begin{figure}[htb]
\begin{center}
{\includegraphics[angle=0,width=15cm]{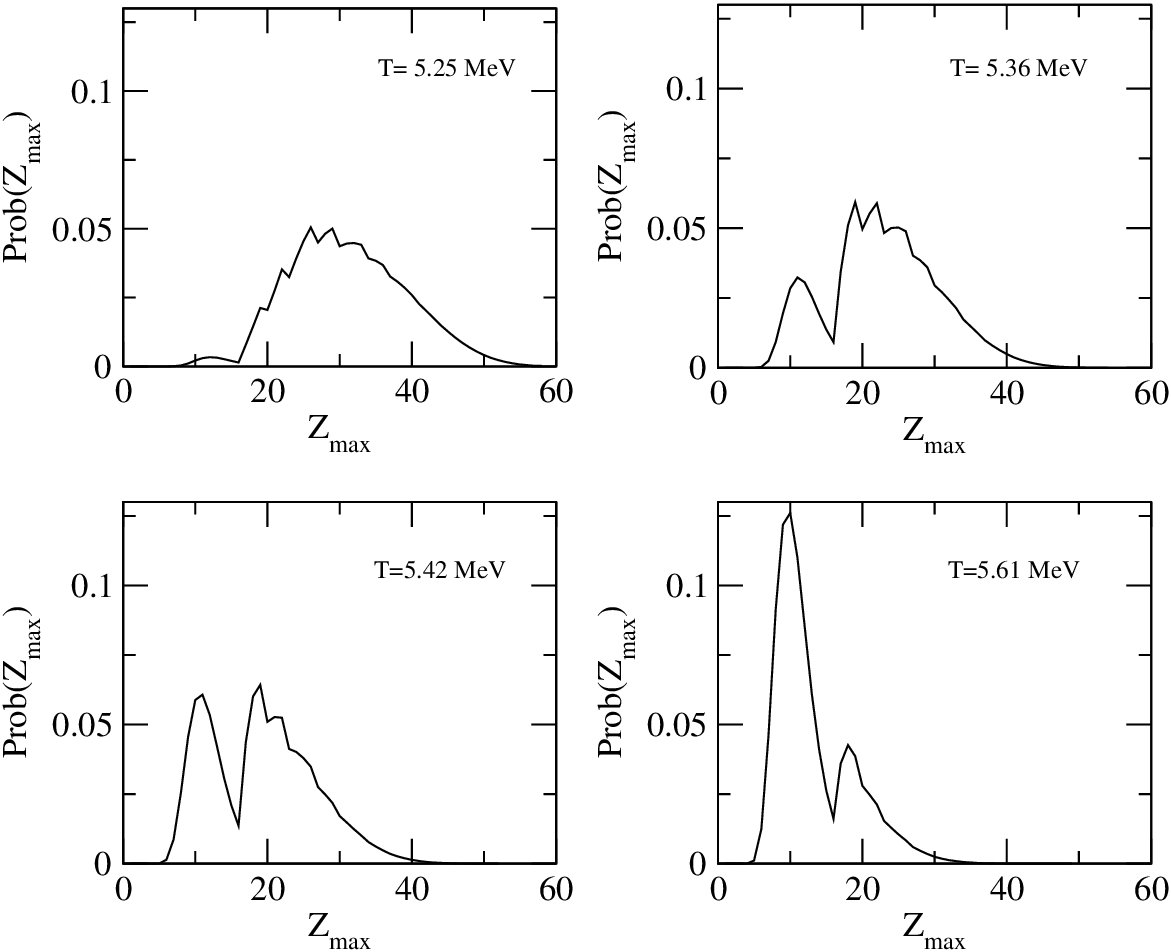}}
\end{center}
\caption{Theoretical values of the probablity of occurrence of a 
$Z_{max}$ plotted as a function of $Z_{max}$ (see eq.(8)).
In a small window of $E^*$ (equivalently $T$ as shown here) a bimodal
distribution is seen.  Similar plot of probability of $A_{max}$ against
$A_{max}$ also shows bimodality.  The plots here use set 1}

\end{figure}


\begin{references}

\bibitem{Bondorf} J. P. Bondorf, A. S. Botvina, A. S. Iljinov, I. N. 
Mishustin and K. Sneppen, Phys. Rep. {\bf 257}, 133 (1995).

\bibitem{Gross} D. H. Gross, Phys. Rep. {\bf 279}, 119 (1997).

 
\bibitem{Randrup} J. Randrup, and S. E. Koonin, Nucl. Phys. A{\bf 471},
355c (1987).

\bibitem{Dasgupta1} S. Das Gupta and A. Z. Mekjian, Phys. Rev. C{\bf 57},
1361 (1998).

\bibitem{Das} C. B. Das, S. Das Gupta, W. G. Lynch, A. Z. Mekjian,
and M. B. Tsang, Phys. Rep. {\bf 406}, 1 (2005).

\bibitem{EOS1} J. B. Elliott et al., Phys. Rev. C {\bf 67}, 024609 (2003).

\bibitem{EOS2} J. A. Hauger et al., Phys. Rev. C {\bf 62}, 024616 (2000).

\bibitem{EOS3} J. A. Hauger et al., Phys. Rev. C {\bf 57}, 764 (1998).

\bibitem{Jennings} B. K. Jennings and S. Das Gupta, Phys. Rev. C
{\bf 62}, 014901 (2000).

\bibitem{Tsang} M. B. Tsang et al., Phys. Rev. C {\bf 64}, 054615 (2001).

\bibitem{Gulminelli} F. Gulminelli and Ph. Chomaz, Phys. Rev. C
{\bf 71}, 054607 (2005).

\bibitem{Pichon} M. Pichon et al., Nucl. Phys. A {\bf 779}, 267 (2006).

\bibitem{Chaudhuri} G. Chaudhuri, S. Das Gupta, and M. Sutton, Phys. Rev. B {\bf 74}, 174106 (2006).

\bibitem{Krishnamachari} B. Krishnamachari, J. McLean, B. Cooper, and J. Sethna, Phys. Rev. B {\bf 54}, 8899 (1996).

\bibitem{Pleimling} M. Pleimling and W. Selke, J. Phys. A: Math. Gen. {\bf 33}, L199 (2000).

\bibitem{Binder} K. Binder, Physica A {\bf 319}, 99 (2003); M. Biskup et al, Physica A {\bf 327}, 583 (2003)
 


\end{references}
\end{document}